\documentstyle[12pt,twoside,fleqn,aggwu,epsfig]{article}

\newcommand{\AmS}{{\protect\the\textfont2
  A\kern-.1667em\lower.5ex\hbox{M}\kern-.125emS}}
\def\Journal#1#2#3#4{{#1} {\bf #2}, #3 (#4)}

\def\EPJ{{\em Eur. Phys. J.} }
\def\NP{{\em Nucl. Phys.} }
\def\NC{{\em Nuovo Cim.} }

\def\PL{{\em Phys. Lett.} }
\def\PR{{\em Phys. Rev.} }

\def\PREP{{\em Phys.  Rep.} }

\def\RMP{{\em Rev. Mod. Phys.} }

\newcommand{\mathbold}[1]{\mbox{\protect\boldmath $\displaystyle #1$}}

\def\rmi{{\rm i}}

\def\intm1p1{{\int\limits_{-1}^{1}}}

\newcommand{\be}{\begin{equation}}
\newcommand{\ee}{\end{equation}}
\newcommand{\bea}{\begin{eqnarray}}
\newcommand{\eea}{\end{eqnarray}}

\newcommand{\real}{{\rm Re}}
\newcommand{\imag}{{\rm Im}}

\newcommand{\bsl}[1]{#1 \!\!\! /}

\title{Nucleon Resonances in Meson Nucleon Scattering with Strangeness Production}

\author{A. Waluyo \address{Center for Nuclear Studies, Department 
        of Physics, The George Washington University, \\ 
        ~\,Washington, D.C. 20052, USA}\thanks{Supported in part by
        US DOE with grant no. DE-FG02-95ER-40907}, C. Bennhold$^{\rm a \ast}$, 
        H. Haberzettl$^{\rm a \ast}$, 
        G. Penner \address{Institut f$\ddot{u}$r Theoretische Physik, Universit$\ddot{a}$t 
        Giessen, D-35392 Giessen, Germany.}\thanks{Supported by DFG and GSI Darmstadt (Germany)}, 
        U. Mosel$^{\rm b\dagger}$, and
        T. Mart\address{Jurusan Fisika, FMIPA, Universitas Indonesia,
        Depok 16424, Indonesia}\thanks{Supported in part by the 
        University Research for Graduate Education
        (URGE) grant.}}

\begin{document}

\maketitle

\begin{abstract}
An effective Lagrangian model in a coupled channels framework is applied to extract nucleon resonance 
parameters. In the K-matrix approximation, we simultaneously analyze all the available data for the transitions 
from $\pi N$ to five possible meson-baryon final states, $\pi N, \pi\pi N, \eta N , K \Lambda$, and $K\Sigma$, 
in the energy range from $\pi N$ threshold up to $\sqrt s = 2$ GeV. 
In this work, we focus our efforts on the $K \Sigma$ channel. In particular, 
we include a set of $\Delta$ resonances 
around $1900$ MeV: the $S_{31}(1900), P_{31}(1910),$ and $P_{33}(1920)$ states. 
In most cases the parameters of nucleon resonances determined by the fits are 
consistent with the commonly accepted values in the literature. We find that in the low-energy region, 
the $S_{11}(1650), P_{11}(1710)$, and 
$P_{13}(1720)$ states are important for the $p(\pi^-,K^0)\Sigma^0$ and $p(\pi^-,K^+)\Sigma^-$ processes
while the $P_{33}(1600)$ and $D_{33}(1700)$ states are 
significant for the pure I=3/2 reaction $p(\pi^+,K^+)\Sigma^+$. 
In the high energy region, we find a $D_{13}$ state at 1945 MeV which is important
for the $p(\pi^-,K^0)\Sigma^0$ reaction while
 the $\Delta$ resonances around $1900$ MeV show up significantly in the
 $p(\pi^-,K^+)\Sigma^-$ and  $p(\pi^+,K^+)\Sigma^+$ processes.
\end{abstract}

\section{INTRODUCTION}
The study of nucleon resonances continues to challenge the field of hadronic physics. 
The interest in this field has grown significantly in the last few years because of new data from 
experimental facilities such as Jefferson Lab, ELSA, MAMI, and Brookhaven National Laboratory.

Nucleon resonances are usually identified in $\pi N$ partial waves analyses, where they are 
labeled by the approximate mass and the $\pi N$ quantum numbers, 
which are the relative orbital angular momentum $L$, the total isospin $T$ and the total angular momentum $J$. 
For example, the $D_{13}(1520)$ resonance has a mass of about $1520$ MeV, isospin $T = 1/2$, total angular momentum 
$J = 3/2$ and decays into an $L = 2$ $\pi N$ state.

There are about 20 $N^\ast (T = 1/2)$ and 20 $\Delta (T=3/2)$ resonance states 
in the 1998 PDG publication \cite{pdg98}. 
Some of these have well established properties, while for others there are still large discrepancies between 
different analyses. One example is the $S_{11}(1535)$ state. Even though this state 
has been given a 4-star label from PDG \cite{pdg98} 
the extracted total and partial widths of this state differ widely
for different analyses \cite{dvl99,tf98,cutkosky80}. 

The study of nucleon resonances generally takes place in two theoretical arenas. There are model studies
using the assumption that nucleons and their excited states are composed primarily of
three valence quarks \cite{NRQM,capstick94,capstick98}. In recent years, 
lattice QCD calculations have been extended to predict masses 
of the low-lying baryons \cite{lee98}.
Although lattice QCD calculations are powerful to determine mass spectra they are more 
difficult to apply to decay widths. 
In order to provide a link between the new and improved data on the one side and the results from lattice QCD and 
quark models on the 
other side, dynamical descriptions using hadronic degrees of freedom are required 
to analyze the data in the various 
asymptotic reaction channels, like $\gamma N, \pi N, \pi\pi N, \eta N, K\Lambda, K\Sigma$, and others.

Clearly, the principal problem faced in this undertaking is the number of open channels. 
With more than one open channel in a 
meson production reaction the unknown couplings, masses and widths of the resonances can only be 
obtained if all open channels are treated simultaneously, thus maintaining unitarity. 

In order to study the structure of nucleon resonances in the higher energy region, one needs to include 
strangeness production channels like $K\Lambda$ and $K\Sigma$. 
Usually models that study these channels \cite{gab2} neglect the hadronic final state interaction, 
performing calculations at tree level. Clearly, this leads to a violation of unitarity. Unitarity can only be 
maintained dynamically if we solve the coupled channels system including all possible channels, 
such as $\eta N \rightarrow K\Lambda$, for which no experimental data are available.

In this study we extend the model of Feuster and Mosel \cite{tf98} by including the previously neglected
$K \Sigma$ final state. We furthermore extend the energy range to $\sqrt s \le 2.0$ GeV, thus including
several additional resonanes between $ 1.9$ GeV $\le \sqrt s \le 2.0$ GeV.  
All $2\pi$ final states like $\rho N$ and $\Delta \pi$ continue to be parametrized
through the coupling to a scalar, isovector $\zeta$-meson with mass $m_\zeta = 2m_\pi$.

\section{Model for Meson Nucleon Scattering}
In general, the scattering amplitude of meson nucleon scattering can be written in terms of a Bethe-Salpeter equation

\bea
T_{\varphi,\varphi}= V_{\varphi,\varphi}(k',q,P) + i \int \frac{d^{4}k"}{(2\pi)^4}
V_{\varphi,\varphi}(k',k",P)G_{BS}(k",P)T_{\varphi,\varphi}(k",q,P)
\label{bseq}
\eea
where $\varphi$ stands for $\pi, \eta, \pi\pi,$ or $K$, and $V_{\varphi\varphi}(k',q,P)$ is the driving potential. 
The four momenta for the incoming, outgoing, 
and intermediate nucleons are $p, p',$ and $p"$, of the outgoing and intermediate mesons are $k'$ and $k"$, and 
of the incoming meson is $q$, so that $P = p + q = p' + k' = p" + k"$ is the total four-momentum. $G_{BS}(k,P)$ 
is the full Feynman meson-baryon propagator and is defined as follows

\bea
G_{BS}(k,P) = \frac {m + \bsl P - \bsl k}{(\mu^2-k^2-i\epsilon)(m^2-(P-k)^2-i\epsilon)}.
\label{bsprop}
\eea

In principle, one needs to solve the Bethe-Salpeter equation in its four-dimensional integral form. 
In practice, approximations are used to obtain solutions.

\subsection{The K-matrix approximation}
The Bethe-Salpeter equation can always be decomposed into two equations:
\bea
K &=& V + V \real (G_{BS}) K, \nonumber \\
T &=& K - \rmi K \imag (G_{BS}) T.
\label{kteqn}
\eea
If we choose to put the intermediate particle on-shell, i.e. neglecting 
the real part of $G_{BS}$, we get 

\bea
i G_{\rm K} &=& -2i(2\pi)^2m_N\delta(k_N^2-m_N^2)\delta(k_m^2-m_m^2)
            \times ~\theta (k_N^0)\theta (k_m^0)(k\!\!\!/_N +m_N) ~
\label{ktprop}
\eea
from which we obtain $K = V$ and thus the $K$-matrix Born approximation.
\be
\left [ T_K \right ] = \left [ V + \rmi V T_K \right ] = 
\left [ \frac{V}{1 - \rmi V} \right ],
\label{tmat}
\ee
where the brackets denote that we have to deal with matrices containing all allowed final state combinations.

\subsection{Background contributions}

Our model contains contributions from the standard Born terms, the $t-$channel exchange via $\rho, a_0,$ and $K^\ast$ and 
the resonance contributions in the $s-$ and $u-$channels. 
The potential $V$ is calculated from the interaction Lagrangian below. 
For the Born and $t$-channel couplings we have:
            
\bea
{\cal L}_{NR} = &-& \frac{g_{\varphi NN}}{2 m_N} \bar N \gamma_5
\gamma_{\mu} (\partial^{\mu} \varphi) N
- g_{sNN} s (\bar N N)
- g_{s\varphi \varphi} s (\varphi^* \varphi) \nonumber \\
&-& g_{vNN} \bar N \left ( \gamma_{\mu} v^{\mu} - \kappa_v
  \frac{\sigma_{\mu \nu}}{4 m_N} v^{\mu \nu} \right ) N -
g_{v\varphi\varphi} \left [ \varphi \times (\partial_{\mu} \varphi)
\right ] v^{\mu},
\label{backcoupl}
\eea
$\varphi$ denotes the asymptotic mesons $\pi, \eta,$ and $K$. A coupling to the $\zeta$-meson is not taken into account. 
$s$ and $v$ are the intermediate scalar and vector mesons ($a_0, \rho, K_0^\ast$ and $K^\ast$) and $v^{\mu \nu} = \partial^{\nu} 
v^{\mu} - \partial^{\mu} v^{\nu}$ is the field tensor of the vector mesons. $N$ is either a nucleon, $\Lambda$, or a $\Sigma$ 
spinor. For $I = 1$ mesons ($\pi$ and $\rho$) $\varphi$ and $v^\mu$ need to be replaced by 
\mathbold {\tau \cdot \varphi} and $\mathbold {\tau \cdot v}^{\mu}$ in the $\varphi, v NN$-couplings.

\section{Results of the Fits to Hadronic Data}

We performed a fit to all hadronic data for the reactions 
$\pi N \to \pi N$, $\pi N \to \pi \pi N$,
$\pi N \to \eta N$, $\pi N \to K \Lambda$, and  $\pi N \to K \Sigma$
up to an energy of W = 2.0 GeV. The results of the Born couplings are
displayed in table~\ref{tab:born}. We point out that the $g_{K \Lambda N}$ and
$g_{K \Sigma N}$ couplings are in accordance with approximate SU(3), in contrast to
ref. \cite{tf98} which found a much smaller $g_{K \Lambda N}$ coupling constant.
In addition to ref. \cite{tf98}
we allow contributions from the $S_{31}(1900), P_{31}(1910)$ and $P_{33}(1920)$ $\Delta$ resonances.
Our best fit results in three  $D_{13}$ excitations, the well-known $D_{13}$(1520) state, 
the 3-star $D_{13}$(1700) state and a new $D_{13}$ state at 1945 MeV with a large width
of 583 MeV.  As discussed in detail in refs. \cite{mart99,benn2000}, it is at present
not clear if this $D_{13}$ state corresponds to the 2-star resonance $D_{13}$(2080) 
listed in the Particle Data Table.  A closer examination of the literature reveals
that there has been some evidence for two resonances in this partial wave
between 1800 and 2200 MeV \cite{cutkosky80}; one with a mass centered around 
1900 MeV and another with mass around 2080 MeV. It is the former
which has been seen prominently in two separate $p(\pi^-, K^0)\Lambda$
analyses \cite{saxon,bell}. 

\begin{table}[!bp]
\caption{Extracted resonance parameters. $^a$: The coupling constant is given instead of the partial width.}
\begin{center}
\begin{tabular}{c||c|c|c|c|c|c|c|c|c|c|c|c}
\hline
& $M$ & $\Gamma_{tot}$ & \multicolumn{2}{c|}{$\Gamma_{\pi N}$} & \multicolumn{2}{c|}{$\Gamma_{\zeta N}$} &
  \multicolumn{2}{c|}{$\Gamma_{\eta N}$} & \multicolumn{2}{c|}{$\Gamma_{K\Lambda}$} & 
   \multicolumn{2}{c}{$\Gamma_{K\Sigma}$}\\
\hline
$L_{2I,2S}$  &[GeV]  & [MeV]&  [MeV] & \% & [MeV] & \% &  [MeV] & \% &  [MeV] & \% &  [MeV] & \% \\
\hline
$S_{11}(1535)$& 1.547 & 244 & 89   & 36 & 9   & 4 & 146& 60& -& -& -  & -  \\
$S_{11}(1650)$& 1.689 & 341 & 189  & 55 & 71  & 22& 1& 0& 80& 23& 0.98$^a$& 0 \\
\hline
$P_{11}(1440)$& 1.473 & 434 & 282  & 65 & 152 & 35& 3.84$^a$& 0& -& -& -& - \\
$P_{11}(1710)$& 1.716 & 235 & 0.43 & 0  & 76  & 32& 54& 23& 73& 31 & 32& 14\\
\hline
$P_{13}(1720)$& 1.736 & 137 & 33   & 25 & 77  & 56& 18& 13& 2& 1& 7& 5 \\
$P_{13}(1900)$& 1.951 & 300 & 80   & 27 & 213 & 71& 0& 0& 5& 2& 2& 0 \\
\hline
$D_{13}(1520)$& 1.507 & 93  & 51   & 55 & 42  & 45& 0.003& 0& -& -& -& - \\
$D_{13}(1700)$& 1.706 & 86  & 0.52 & 0  & 33  & 38& 42& 49& 10& 12& 1& 1 \\
$D_{13}(1900)$& 1.945 & 583 & 108  & 19 & 439 & 75& 28& 5& 8& 1& 0.08& 0 \\
\hline
\hline
$S_{31}(1620)$& 1.616 & 148 & 56   & 38 & 92  & 62& -& -& -& -& -& - \\
$S_{31}(1900)$& 1.938 & 334 & 121  & 36 & 166 & 50& -& -& -& - & 47& 14\\
\hline
$P_{31}(1910)$& 1.967 & 605 & 119  & 20 & 452 & 74& -& -& -& -& 34& 6 \\
\hline
$P_{33}(1232)$& 1.229 & 110 & 110  & 100&  -  & -  & - & - & - & - & - & - \\
$P_{33}(1600)$& 1.677 & 296 & 44   & 15 & 252 & 85 & - & - & - & - & 0.59$^a$& 0  \\
$P_{33}(1920)$& 2.077 & 689 & 114  & 17 & 575 & 83 & - & - & - & - & 0.60  & 0 \\
\hline
$D_{33}(1700)$& 1.672 & 583 & 73   & 13 & 510 & 87 & - & - & - & - & 0.31$^a$& 0 \\
\hline
\end{tabular}
\end{center}
\label{tab:res}
\end{table}

\begin{table}[!bp]
\caption{Couplings of the meson nucleon as obtained in the fits.}
\begin{center}
\begin{tabular}{c|r||c|r||c|r}
\hline
g              &  Value & g                     & Value & $\kappa$                  & Value \\
\hline
\hline
$g_{\pi NN}$   &  13.21 & $g_{\rho NN}$         &  1.95 & $\kappa_{\rho NN}$        &  2.29 \\
\hline
$g_{KN\Sigma}$ &   1.41 & $g_{K^\ast N\Sigma}$  & -1.12 & $\kappa_{K^\ast N\Sigma}$ & -0.38 \\
\hline
$g_{\eta NN}$  &   0.36 & $g_{a_{0}NN}$         &  0.98 &  -                        &    -  \\
\hline
$g_{KN\Lambda}$& -13.98 & $g_{K^\ast N\Lambda}$ & -4.96 & $\kappa_{K^\ast N\Lambda}$& -0.37  \\
\hline
\end{tabular}
\end{center}
\label{tab:born}
\end{table}

We now focus our discussion on the role of background versus nucleon resonant contributions in the $K\Sigma$ channel 
at different energies.
All resonance parameters we have extracted are given in table~\ref{tab:res} which
shows that nucleon resonances with the 
largest decay width into the $K\Sigma$ channel are the $P_{11}(1710), P_{13}(1720), S_{31}(1900),$ 
and $P_{31}(1910)$ states. 

\begin{figure}[hbt]
\epsfig{file=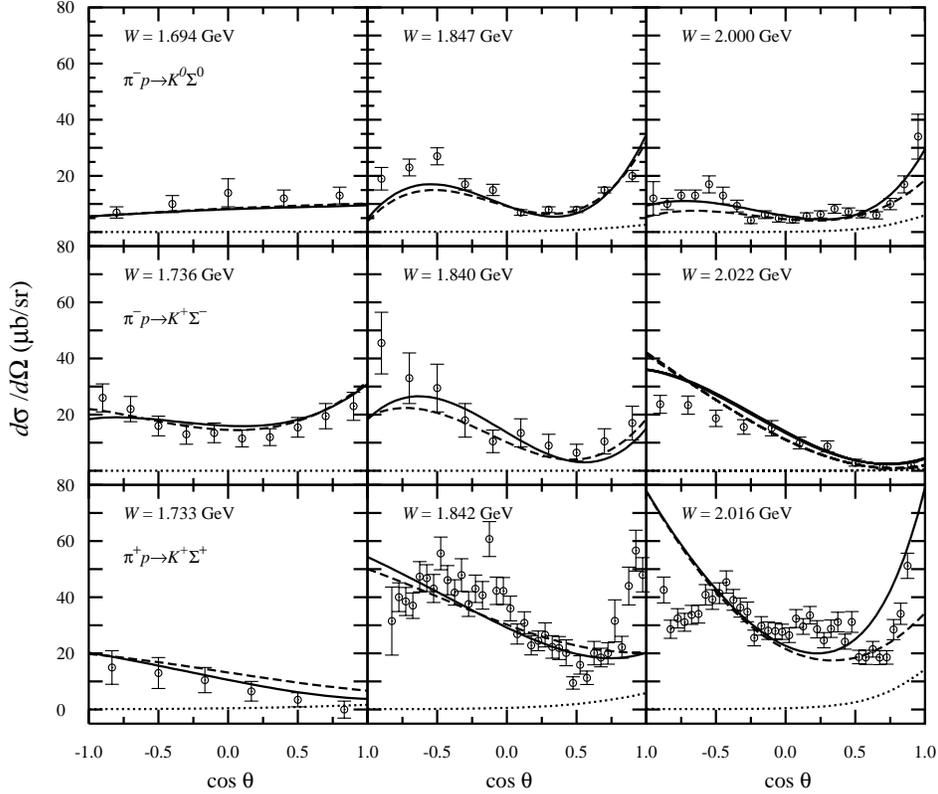,height=110mm}
\vspace{-1cm}
\caption{Differential cross sections of the three $\pi N \rightarrow K \Sigma$ reactions. 
Shown are the results from the full calculation (solid line), 
Born contributions only (dotted line), and the full calculation without K* (dashed line).
The data are from ref. \cite{Ksigmadata}}
\label{fig:born}
\end{figure}

Figure~\ref{fig:born} shows that the structure of the differential cross section for $\pi N \rightarrow K\Sigma$ 
is dominated by resonances. At higher energies we find forward peaking behavior from the $t-$channel $K^\ast$ exchange. 
This can be seen especially in the $p(\pi^+,K^+)\Sigma^+$ reaction. This confirms what has been found 
in reference \cite{tf98} about the role  of vector meson $t-$channel exchange at higher energies.

\begin{figure}[hbt]
\epsfig{file=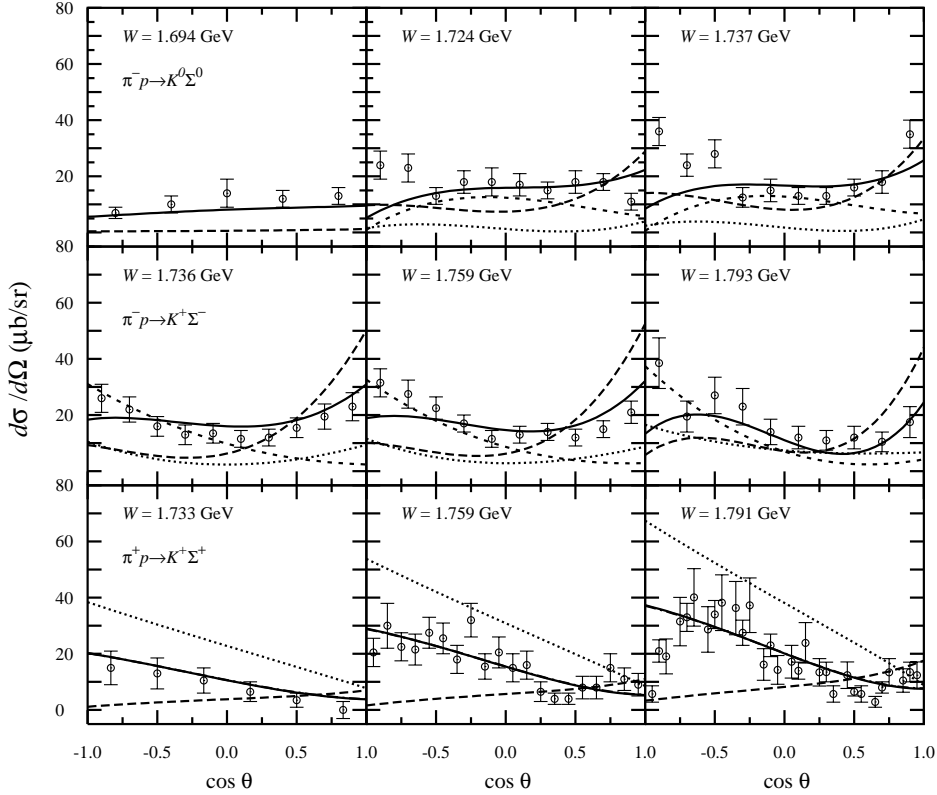,height=110mm}
\vspace{-1cm}
\caption{ Differential cross sections of the three $K\Sigma$ reactions in 
the low energy region. Shown for the $p(\pi^-,K^0)\Sigma^0$ and $p(\pi^-,K^+)\Sigma^-$ reactions are the 
full calculation (solid line), without the $P_{11}(1710)$ (long-dashed line), without the $P_{13}(1720)$ 
(short-dashed line), and excluding both the $P_{11}(1710)$ and 
$P_{13}(1720)$ at the same time (dotted line).
For $p(\pi^-,K^0)\Sigma^0$ at W = 1.694 GeV we only show the full calculation (solid line) and 
without the $S_{11}(1650)$ (dashed line). 
For the I=3/2 reaction $p(\pi^+,K^+)\Sigma^+$ we show
the full calculation (solid line), without the $P_{33}(1600)$ (dashed line) and without the $D_{33}(1700)$ (dotted line).}
\label{fig:low}
\end{figure}

In fig. ~\ref{fig:low} we examine the role of nucleon resonances in the low energy region.
Since the extracted partial widths for $P_{11}(1710)$ and $P_{13}(1720)$ into the $K\Sigma$ channel are $32$ and $73$ MeV, 
one expects these resonances to have an important contribution in this energy region. 
For the $K^0 \Sigma^0$ and $K^+ \Sigma^-$ reactions, we find that the main contribution at low energies indeed comes from these 
two resonances. 
For the $p(\pi^-,K^+)\Sigma^-$ reaction, it can be seen clearly that the $P_{11}(1710)$ 
gives backward peaking behavior 
while the $P_{13}(1720)$ is responsible for forward peaking behavior.
The role of the $P_{11}(1710)$ is consistent with recent studies 
of kaon photoproduction \cite{cb99} and of $K\Sigma$ production in $NN$-scattering \cite{sib98}, $NN \rightarrow NK\Sigma$, 
where the $P_{11}(1710)$ state 
was identified as a major contribution. Figure~\ref{fig:low} also demonstrates
that very close to threshold ($W_{threshold} = 1690$ MeV for $K^0 \Sigma^0$ production)
at $W_{cm} = 1694$ MeV, the structure of the differential cross section of the $p(\pi^-,K^0)\Sigma^0$ reaction
comes mainly from the $S_{11}(1650)$ state. 
The extracted mass of this resonance is $1689$ MeV, the total width is $341$ MeV and the coupling 
constant to the $K\Sigma$ channel is $0.98$. Even though this resonance lies below the threshold of the $p(\pi^-,K^0)\Sigma^0$ 
reaction, it is clearly its proximity to the $K^0 \Sigma^0$ production threshold that leads
to its dominating influence on the threshold cross section.

Now we come to the most interesting reaction in the $K\Sigma$ channel, which is $p(\pi^+,K^+)\Sigma^+$. 
This reaction is a pure isospin $3/2$ process. Therefore, as expected, we do not see any effects of 
the $P_{11}(1710)$ and $P_{13}(1720)$ states.
Fig. ~\ref{fig:low} shows that the effects of the $P_{33}(1600)$ and $D_{33}(1700)$ states are
large in the low-energy region.
The extracted masses for the $P_{33}(1600)$ and $D_{33}(1700)$ are $1677$ MeV and $1672$ MeV, respectively, 
but their total widths are $296$ MeV and $583$ MeV which leads to broad resonance tails and 
significant influence on 
the cross section.  
From fig. ~\ref{fig:low} it is clear that the $D_{33}(1700)$ gives 
the largest contribution which interferes destructively 
with a smaller signal from the $P_{33}(1600)$.

\begin{figure}[hbt]
\epsfig{file=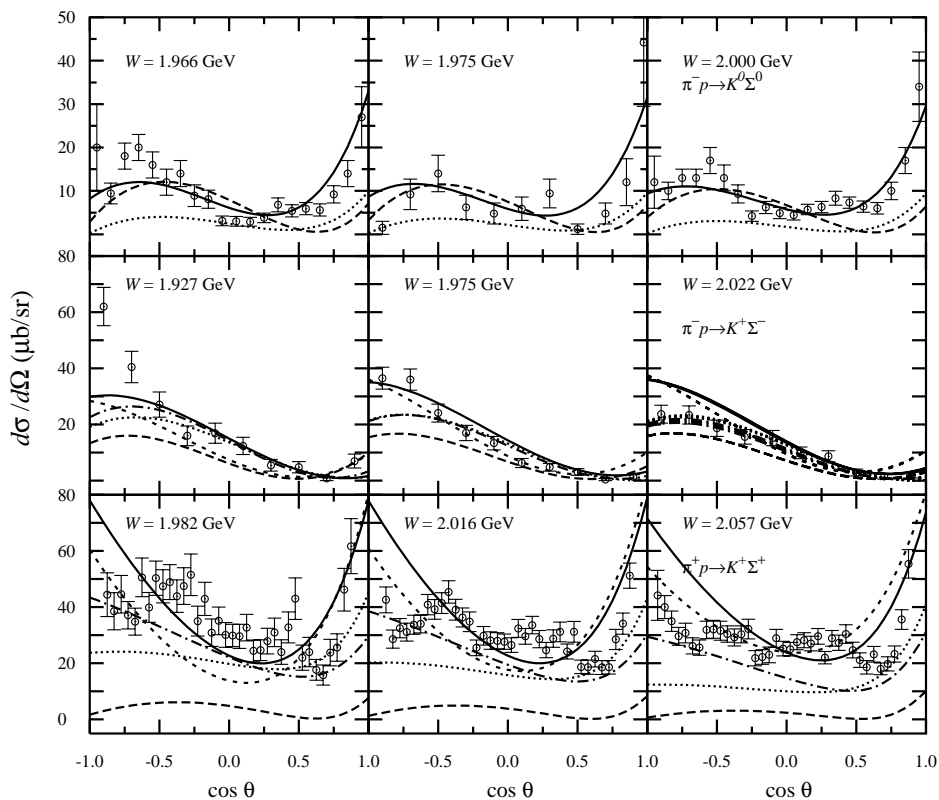,height=110mm}
\vspace{-1cm}
\caption{Differential cross sections of the three $K\Sigma$ reactions at
higher energies. For all three reactions we display the full calculation (solid line) 
and results leaving out the $S_{31}(1900), P_{31}(1910)$,
and $P_{33}(1920)$ states (long-dashed line). 
Furthermore, for the $p(\pi^-,K^0)\Sigma^0$ reaction we also show calculations without the 
$P_{13}(1900)$ (short-dashed line), and without the $D_{13}(1900)$ (dotted line). 
Finally, for the $p(\pi^-,K^+)\Sigma^-$ 
and $p(\pi^+,K^+)\Sigma^+$ reactions, calculations are displayed 
without the $S_{31}(1900)$ (short-dashed line), 
without the $P_{31}(1910)$ (dotted line) and without the $P_{33}(1920)$ (dot-dashed line).}
\label{fig:high}
\end{figure}

In fig. ~\ref{fig:high} the $K\Sigma$ channel is studied at higher energies.
The three $\Delta$ states,  the $S_{31}(1900), P_{31}(1910)$, and $P_{33}(1920)$ are important in 
both isospin $1/2$ and $3/2$ channels.




For the $K^0\Sigma^0$ reaction, 
we find that the $S_{31}(1900), P_{31}(1910),$ and $P_{33}(1920)$ contribute
only at forward angles. At backward angles 
we find that the main contribution to the cross section of this reaction comes from the $D_{13}(1900)$.
For the $K^+\Sigma^-$ reaction, we find that the $S_{31}(1900)$ 
and $P_{31}(1910)$ supply the main contributions to the structure of the cross section.  For the $K^+\Sigma^+$ 
reaction, all three $\Delta$ resonances build up the cross section. 
Even though the $P_{33}(1920)$
has only a small partial width into this channel ($\Gamma_{K\Sigma} = 593$ keV), its role is 
significant in the backward angle region. 
The main contribution to the cross section comes from the $S_{31}(1900)$ and $P_{31}(1910)$.
However, the model cannot reproduce the structure of the differential 
cross section data. This structure might come from high-spin $N^\ast$ resonance contributions (such as spin $5/2$ or $7/2$) 
or different background mechanisms not included in the model.

\section{Conclusion}
We  have investigated the role of nucleon resonances within a coupled channels framework. 
In the $K-$matrix approximation and with five allowed asymptotic states, $\pi N, \pi\pi N,$ $\eta N,
K\Lambda,$ and $K\Sigma$, we fit the hadronic data in the energy range $\pi N$ threshold up to $W_{cm} = 2$ GeV.

Focussing on the $K \Sigma$ channel we find that the structure of the differential cross section comes 
purely from resonance contributions. We also find that $P_{11}(1710)$ and $P_{13}(1720)$ give the main 
contributions to the cross section in the low energy region of the $p(\pi^-,K^0)\Sigma^0$ and $p(\pi^-,K^+)\Sigma^-$ 
reactions. At higher energies, the $D_{13}(1900)$ dominates the $p(\pi^-,K^0)\Sigma^0$ reaction while 
the $S_{31}(1900)$ and $P_{31}(1900)$ 
dominate the cross section for the $p(\pi^-,K^+)\Sigma^-$ reaction. For the $p(\pi^+,K^+)\Sigma^+$ reaction, 
we find no satisfactory mechanism in the high energy region, even though
the $S_{31}(1900)$ and $P_{31}(1900)$ states play an important role in this regime. 
As the next step the photoproduction reaction $\gamma N \rightarrow K \Sigma$ will be studied
along with the hadronic production reactions discussed here.

This work was supported by DOE grant DE-FG02-95ER-40907 (AW, CB, and HH), DFG and GSI Darmstadt 
(GP and UM), and a University Research for Graduate Education (URGE) grant (TM).

\end{document}